\documentclass{emulateapj}

\begin{document}

\shorttitle{The origin of remnants in Virgo}
\shortauthors{Mart\'{i}nez-Delgado et al.}

\title{The Virgo stellar over-density: Mapping the infall of the
Sagittarius tidal stream onto the Milky Way disk
}

\author{David Mart\'{i}nez-Delgado\altaffilmark{1,2},
Jorge Pe\~{n}arrubia\altaffilmark{3,4}, Mario Juri\'c\altaffilmark{5,6},  Emilio J.
Alfaro\altaffilmark{2}, Zeljko Ivezi\'c\altaffilmark{5}}

\altaffiltext{1} {Instituto de Astrof\'isica de Canarias, La Laguna, Spain}
\altaffiltext{2} {Instituto de Astrof\'isica de Andaluc\'\i a (CSIC), Granada, Spain}
\altaffiltext{3} {Max-Planck Institute fuer Astronomie, Heidelberg, Germany}
 \altaffiltext{4} {Department of Physics and Astronomy, University of Victoria, Victoria BC, V8P1A1, Canada}
\altaffiltext{5} {Department of Astrophysical Sciences, Princeton, NJ 08544 USA}\altaffiltext{6} {Institute for Advanced Study, School of Natural Sciences, Einstein Drive, Princeton, NJ 08540}

\begin{abstract}

The recently discovered Virgo stellar over-density, which
expands over $\sim$1000deg$^2$ perpendicularly to the Galactic disk plane (7$< Z <$ 15 kpc, R$\sim$ 7 kpc), is the largest clump of tidal debris ever detected in the outer halo and is likely related with the accretion of a nearby dwarf galaxy by the Milky Way. We carry out  N-body simulations of the Sagittarius stream to show that this giant stellar over-density is a confirmation of theoretical model predictions for the leading tail of the Sagittarius stream to cross the Milky Way plane in the Solar neighborhood. Radial velocity measurements are needed to confirm this association and to further constrain the shape of the Milky Way dark matter halo through a new generation of theoretical models. If the identification of Virgo over-density and the Sagittarius leading arm is correct, we predict highly negative radial velocities for the stars of Virgo over-density. The detection of this new portion of the Sagittarius tidal stream would represent an excellent target for the on-going and future kinematic surveys and for dark matter direct detection experiments in the proximity of the Sun.

\end{abstract}


\section{Introduction}\label{int}

In the last decade, various large scale CCD surveys (Sloan Digital Sky
Survey, 2MASS\footnote{Two Micron All Sky Survey}, QUEST) have for the first
time found strong evidence for the presence of a significant amount of substructure in the halo of the Milky Way  in form of long tidal streams or stellar clumps (Newberg et al. 2002; Majewski et al. 2003). These have been
 interpreted as the remnants of latest mergers occurring as a part of the hierarchical build-up of our Galaxy. In addition, N-body simulations of these merger events (Helmi 2004; Mart\'inez-Delgado et al.  2004a; Law, Johnston \&
Majewski 2005; Pe\~narrubia et al. 2006) have been extensively used to, firstly, associate the tidal debris 
with their progenitor galaxies, secondly, determine the dynamical history of the progenitor satellites and, thirdly, constrain the distribution of dark matter in the  Milky Way. The ultimate goal of these studies is to infer the formation process of the Milky Way from the properties of the present fossil records.

Much of the observational and theoretical effort has focused specially on the two largest
tidal streams discovered so far: the tidal stream of the Sagittarius (Sgr) dwarf galaxy,
which wraps around the Galaxy in a highly inclined orbit with respect to the Milky Way disk (Ibata et al. 2001; Mart\'\i nez-Delgado et al. 2001; Majewski et al. 2003) and the Monoceros tidal stream (Yanny et al. 2003; Ibata et al. 2003), a low-latitude tidal stream whose progenitor, possibly the controversial Canis Major dwarf galaxy (Martin
et al. 2004; Bellazzini et al. 2004; Mart\'\i nez-Delgado et al. 2005), moves on a prograde, nearly circular orbit (Pe\~narrubia
et al. 2005), according to the recent proper motions measured by Dinescu et al. (2005). 
Additionally, Rocha-Pinto et al.(2004) reported the discovery of a new possible tidal stream in the Triangulum/Andromeda constellations,  although subsequent theoretical simulations identified this tidal debris as a part of  a more distant, metal poor wrap of the Monoceros tidal stream progenitor (Pe\~narrubia et al.
2005; Mart\'\i nez-Delgado et al. 2007, in preparation).

In addition to these two large tidal streams, the analysis of large samples
of stellar tracers of halo substructure (RR Lyrae, F-type turnoff
stars) have revealed the presence  of numerous stellar over-densities possibly associated with
unknown merger events in the outer halo. In the Virgo constellation, several groups have reported
numerous detections of stellar over-densities using different tracers: \\
(i) The most conspicuous one is a diffuse concentration of metal poor, old stars situated at $\sim$
20~kpc from the Sun in the direction of the Virgo constellation, first detected  by means of the
identification of its associated main-sequence turnoff (MSTO) in a color-magnitude diagram (CMD) of
a
SDSS slice (named S297+63-20.0; Newberg et al. 2002). Additionally, Vivas \& Zinn (2003) reported a
clump of 21 RR Lyrae stars located at $\sim$ 20 kpc from the sun in the same region of the sky (see
also Ivezic et al. 2005 and Vivas \& Zinn 2006). This over-density (named the {\it 12$^{h}$.4 clump}) extends over an area
of the sky  of $15^\circ\times 1^\circ$ (corresponding to to a spatial size of $5.3$ kpc$\times 0.3$
kpc at $D\simeq 20$ kpc) and display  a line-of-sight depth of only $\simeq 1.5 $kpc (Vivas 2002),
which is considerably smaller than that of a similar RR Lyrae over-density detected at the apocenter
of the Sgr stream ($\sim$ 4 kpc; Vivas 2002). Owing to its possible tidal origin, they denoted it  as the Virgo stellar stream (VSS). Mart\'inez-Delgado et al. (2004a) suggested
that the VSS might be related to the complex debris structure of the Sgr stream predicted by the
theoretical models in this region of the sky. However, spectroscopic follow-up of a sample of these
RR Lyrae stars by Duffau et al. (2006) showed a fairly narrow velocity distribution that pointed to a progenitor galaxy less massive than the Sgr dwarf and, therefore, to a hitherto unknown tidally disrupted dwarf galaxy. Using SDSS photometric data,  these authors estimated that the VSS covers roughly an area of 106
deg$^{2}$ in the sky.\\

(ii) More recently, Juri\'c et al. (2006) detected  a remarkable excess of main-sequence
stars towards the direction of Virgo constellation, named the {\it Virgo over-density }(VOD). 
These authors construct a three-dimensional density distribution of Milky Way stars
based on the position of $\sim$ 48 million stars obtained  by the SDSS in order to identify stellar over-densities.
  The VOD density  map reveals a giant, diffuse lump of stars without any evidence of a distinct core that 
extends over more than $\sim$ 1000 deg$^{2}$ (see Fig. 1) towards the direction
$(l,b)=(300^\circ,65^\circ)$. This stellar over-density shows an remarkably extended vertical structure perpendicular to the Galactic plane ($\Delta Z \simeq 8$ kpc, with heliocentric distances between 5-17 kpc) that possibly crosses the Galactic plane, extending into the southern Galactic hemisphere (see also Fuchs, Phleps \& Meisenheimer 2006). The density of VOD main-sequence stars peaks at
 $\sim$ 16 kpc from the Sun (or RR Lyrae magnitude of $\sim$ 16.7), in good agreement with
the RR Lyrae distance determination. These authors suggest that the giant stellar over-density
might be related with a hitherto unknown merger event of a nearby, low metallicity
dwarf galaxy .Alternatively, Xu, Deng \& Hu (2006) suggest that this stellar lump could be related to a non-axisymmetric Galactic stellar spheroid (see Sec. ~\ref{triaxial}).

 Although VOD and VSS are coincident in position and display
 a huge sky-projected size, the
different vertical structure suggests that they are independent systems. In absence of radial velocity measurements for the VOD, it is not possible to address whether they share a common origin or are completely unrelated. 
For this reason, we will refer
to them with different names in this paper, following the nomenclature given
to each over-density in the original papers.

The main aim of this paper is to investigate the nature of the
 VOD and to analyze whether this stellar lump could be
 related to the Sgr stream through a comparison between up-to-date theoretical models and the available observational data. The structure, kinematics and possible origin of the VSS will be analyzed
in a companion paper.

In Sec. 2 the theoretical models of Sgr tidal stream are described. In Sec.3 we discuss the possible origins of the VOD. In particular, we explore  the presence of tidal debris from the Sgr tidal stream in Virgo by means of update
theoretical models and compare its expected structural parameter with those derived for Virgo.
Finally, conclusions are given in Sec. 4.

\section{THEORETICAL MODELS}\label{model}

The giant stellar over-density detected in the Virgo constellation is located in a
region of the Milky Way where several theoretical models predict the presence of Sgr galaxy debris.
In order to explore a possible association of the VSS and VOD with the Sgr tidal stream, we shall
compare their derived spatial and kinematic properties with those obtained from up-to-date theoretical models.
 
Our Galaxy model consists of a Miyamoto-Nagai (1975) disk, a Hernquist (1990) bulge and a Navarro,
Frenk \& White (1996) dark matter halo (hereafter, NFW halo).
The gravitational potential of each of those components in cylindrical coordinates is:
\begin{eqnarray}
\Phi_d({\bf r})=-\frac{G M_d}{\sqrt{R^2+(a+\sqrt{z^2+b^2})^2}}, \label{eq:phid} \\
\Phi_b({\bf r})=-\frac{G M_b}{r+c},\label{eq:phib} \\
\label{eq:phih} \Phi_h({\bf r})=\frac{G M_h}{\ln(1+\frac{r_{\rm vir}}{r_s})-\frac{r_{\rm
vir}}{r_s+r_{\rm vir}}}\frac{q_h}{2r_s}\times \\ \nonumber 
\bigg[\int_0^\infty\frac{m(u)}{1+m(u)}\frac{{\rm d}u}{(1+u)\sqrt{q_h^2+u}}-2\bigg].
\end{eqnarray}
There, $r_{\rm vir}, r_s$ are the virial and scale radii, respectively, $M_h=M_h(r_{\rm vir})$ and
$r^2=R^2+z^2$. The potential of an axi-symmetric NFW halo was calculated from Chandrasekhar (1960)
using elliptic coordinates
\begin{eqnarray}\label{eq:mcoord}
m^2(u)=\frac{R^2}{r_s^2(1+u)}+\frac{z^2}{r_s^2(q_h^2+u)},
\end{eqnarray}
where $q_h$ is the axis-ratio of iso-density surfaces. In this contribution, we shall denote {\it
oblate} and {\it prolate} halos as those with $q_h<1$ and $q_h>1$, respectively. For spherical halos
Eq.~(\ref{eq:phih}) reduces to
\begin{eqnarray}
\Phi_h=-\frac{G M_h}{\ln(1+\frac{r_{\rm vir}}{r_s})-\frac{r_{\rm vir}}{r_s+r_{\rm
vir}}}\frac{\ln(1+r/r_s)}{r}\equiv \\ \nonumber
-V_c^2\frac{\ln(1+r/r_s)}{r/r_s}.
\label{eq:phih_sph}
\end{eqnarray}

Following Johnston et al. (1999) we fix the disk and bulge parameters as $M_d=1.0\times
10^{11}M_\odot$, $M_b=3.4 \times 10^{10} M_\odot$, $a=6.5 $kpc, $b=0.26$ kpc and $c=0.7 $kpc. The
Milky Way halo parameters at $z=0$ were taken from Klypin, Zhao \& Somerville (2002) being
$M_h=1.0\times 10^{12}M_\odot$, $r_{\rm vir}=258 $kpc, $r_s=21.5 $kpc, which leads to a
concentration at the present epoch of $c=r_{\rm vir}/r_s=12$.
According to the results of Pe\~narrubia et al. (2006), the evolution of the host Galaxy potential is not reflected in the present properties of tidal streams, so that we use a static Milky Way potential for simplicity.

The actual disruption of satellites is modelled by `live' (i.e. self-consistent, self-gravitating)
N-body realizations of a King (1960) model, with a dimensionless central potential $W_0=4$, or
concentration parameter $c\equiv\log_{10}(r_t/r_k)\simeq 0.84$, where $r_k$ and $r_t$ are the King
and tidal radii, respectively. Our satellite models have $N=10^5$ particles. 
The initial and final (i.e present) masses are $M_s(t_0)=10^9M_\odot$ and $M_s(t_f)=5\times 10^8
M_\odot$, respectively. The initial King and tidal radius are $r_k(t_0)=0.58$ kpc and
$r_t(t_0)=4.01$ kpc.

The equation of motion for each satellite particle is:
\begin{equation}\label{eq:eqmot}
\frac{d^2 {\bf r}_i}{d t^2}=-{\bf \nabla}(\Phi_s+\Phi_d+\Phi_b+\Phi_h)_i
\end{equation}
where  $\Phi_s$ is the self-gravitational potential of the satellite galaxy and $i=1,...,N$.
We use {\sc superbox} (Fellhauer et al. 2000) to calculate $\Phi_s$ at each time-step and solve
Eq.~\ref{eq:eqmot} through a leap-frog scheme with a constant time-step of $\Delta t=0.65$ Myr,
which is about $1/100$th the satellite's dynamical time.\\ 
Note that the Galactic potential is static and, therefore, Eq.~\ref{eq:eqmot} does not implement the
response of the Milky Way to the presence of Sgr. We also neglect the effects of dynamical friction
on the Sgr's orbit (as suggested by Law et al. 2005).

We have reproduced the N-body models presented in Law et al. (2005). These models
reproduce the following observational constraints\footnote{It is also important to remark that we have not input any observational data of the VOD  in the fitting of these simulations, that are only  based in the fitting of an independent set of data of the Sgr tidal stream.}:
\begin{enumerate}
\item The heliocentric position and radial velocity of the Sgr dwarf are $(D,l,b)=(24 {\rm kpc},
5.6^\circ,-14.2^\circ)$ and $v_r=171$ km/s, respectively.
\item The orbital plane of Sgr has an inclination with respect to the Milky Way disk of $i\simeq
76^\circ$.
\item The averaged heliocentric distance for Sgr leading debris is $D\sim 50$ kpc.
\end{enumerate}
Items 1. and 2. above show that we have observational measurements for five of the six coordinates
that determine the Sgr orbit. Law et al. (2005) surveyed the unknown coordinate
(namely, the tangential velocity component, $v_{\rm tan}$) so that the resulting N-body models
reproduced all the observational constraints listed. 

Among other free
parameters,  Law et al. (2005) also explored which halo axis-ratio ($q_{h}$) would
produce the best-fitting model to the available observational data. They
found that, whereas the geometry and kinematics of the trailing arm are
scarcely sensitive to the adopted $q_{h}$, radial velocity measurements of the
leading arm were best matched by prolate ($q_{h}>1$) halo models, in
agreement with Helmi (2004). However, Johnston, Law and Majewski (2005) showed that Sgr models in
prolate halos cannot reproduce the observed precession rate in the youngest pieces of the Sgr
stream. Furthermore, they were able to constrain the halo axis-ratio to be $q_{h}=0.83--0.92$,
excluding models with $q_{h}>1$ at a $3\sigma$ level.
In view of these results, we have performed N-body simulations where the halo axis-ratio was fixed
either to $q_{h}=0.8$ or to $q_{h}=1.4$, excluding for the sake of brevity spherical halos, which
neither reproduce the radial velocity variation along the stream nor its precession rate.

In a  coordinate system where the Sun is located at $(X,Y,Z)=(8,0,0)$ kpc, with a velocity of
$(U,V,W)=(-10, -220, 7)$ km/s (Binney \& Merrifield 1998), the Sgr velocities that best fit the
kinematic and spatial distributions of stream debris are $(237,-35,220)$ km/s for $q_h=0.8$
(oblate halo) and $(244,-39, 249)$ km/s for $q_h=1.4$ (prolate halo).

With those inputs, we have evolved the N-body satellite model from 4 Gyr in the past to the present.

\section{THE ORIGIN OF THE VIRGO OVER-DENSITY }\label{vod}
In this Section we analyse the likelihood of various scenarios that may shed light on the nature of the VOD. 

\subsection{The signature of a tri-axial Galactic stellar  halo}\label{triaxial}

An alternative explanation for the presence of the unexpected MS feature in the CMD  reported in
several studies (and so-called the VOD: Newberg et al. 2002; Duffau et al. 2006; Juric et al. 2006; see Sec.1) is to postulate the existence of a
 non-axisymmetry Galactic model component, such as a triaxial halo (Newberg \& Yanny
2005; Xu et al. 2006). We have explored this possibility by inspecting a sample of deep CMDs (with
limiting magnitude $R_{lim}\sim 24.5$) of Galactic fields taken with the Wide Field Camera (WFC) at the prime focus of Isaac Newton 2.5-meter telescope (La Palma,Spain)  during different observing runs devoted to the search of tidal debris 
associated to known Galactic satellites or globular clusters (Mart\'\i nez-Delgado et al. 2004b).
Table 1 lists the position of our fields with positive and negative detections of the MS feature in
the CMD. These negative detections can be  interpreted as the lack of tidal debris  or that
its surface brightness is extremely low to be detected in the small field of view
of the WFC ($35\arcmin \times 35\arcmin$)\footnote{Mart\'\i nez-Delgado et al. 2004a estimated the
limit SB detection of our CMD technique is $\Sigma \sim 32$ mag arcsec$^{-2}$}.

Fig.~\ref{densitymap}  shows the position of these negative detection (open circles)  over-plotted in
the density map of VOD derived by Juri\'c et al (2006) from SDSS observations.  The most interesting
result is a negative detection $\sim 20\arcdeg$ North of the center of our target field (see Table
1) that clearly shows a decrease  of  stream star counts in a small area of the sky. 
Although our sparse pencil beam survey cannot constrain the spatial extent of 
this stellar population in the sky, the distribution of negative detections strongly indicates that this MS population is not likely to be a manifestation of a non-axisymmetric Galactic model component, but is probably associated with the remnants of a dwarf galaxy under tidal disruption, in
agreement with previous studies.
Juri\'c et al. (2006) have reached the same conclusion by analyzing the vertical (Z) distribution of
VOD debris from their sample of SDSS G stars ($0.10 < r-i < 0.15$).

\subsection{A tidal debris of the Sgr dwarf?}\label{sgr}

\subsubsection{Sky projected density maps}\label{map}

It is well understood both from observations and theoretical modeling
that the Sgr leading arm turns around and falls towards the Galactic plane in the
proximity of the Solar neighborhood, yielding a vertical debris structure that might cross the
Galactic plane and possibly extend into the southern Galactic
hemisphere (Majewski et al. 2003; Mart\'inez-Delgado et al. 2004a; Law et al. 2005). This downward
flow of the Sgr stars onto the Galactic disk may share similar
characteristic to those derived for the VOD in the stripe covered by the SDSS.
As a guide for the reader, we illustrate this  putative scenario  in Fig.~\ref{esquema} together with the identification of the different parts of the Sgr tidal stream obtained from our theoretical models (see Sec. ~\ref{3dmap} for a three-dimensional comparison).

To examine whether the VOD is a part of the
Sgr tidal structure, we compare the predictions from the
 theoretical models outlined in 
Sec.~\ref{model} to the SDSS data presented in Juri\'c et al. (2006).When comparing the theoretical stream models against the SDSS, it is
important to keep in mind that the observational data may be biased by
systematic over or underestimates of the distance scale. The stellar
distances in SDSS data were derived using an assumed photometric paralax
relation whose shape is well constrained but the zero-point may be
offset by as much as 0.4mag (Juri\'c et al. 2006). This would translate
into 20
observed center of VOD to anywhere from 8 to 12 kpc. However, even the
worst case scenario would be compatible with the conclusions that we are
to present.

Secondly, the vertical shape and density profile of VOD (Fig. 4, bottom
left panel) depend strongly on the details of the (still preliminary)
Galactic stellar halo model. The density gradients along this vertical
structure should therefore be interpreted with caution, especially at
low $Z$ (closer to the disk). There, the number density of Galactic
stars dominates over the stars in VOD and even a few percent mismatch in
the Galactic model may lead to an artificial increase (or decrease) of
the residual stellar density ascribed to VOD.

\begin{figure}
\epsscale{1.0}
\plotone{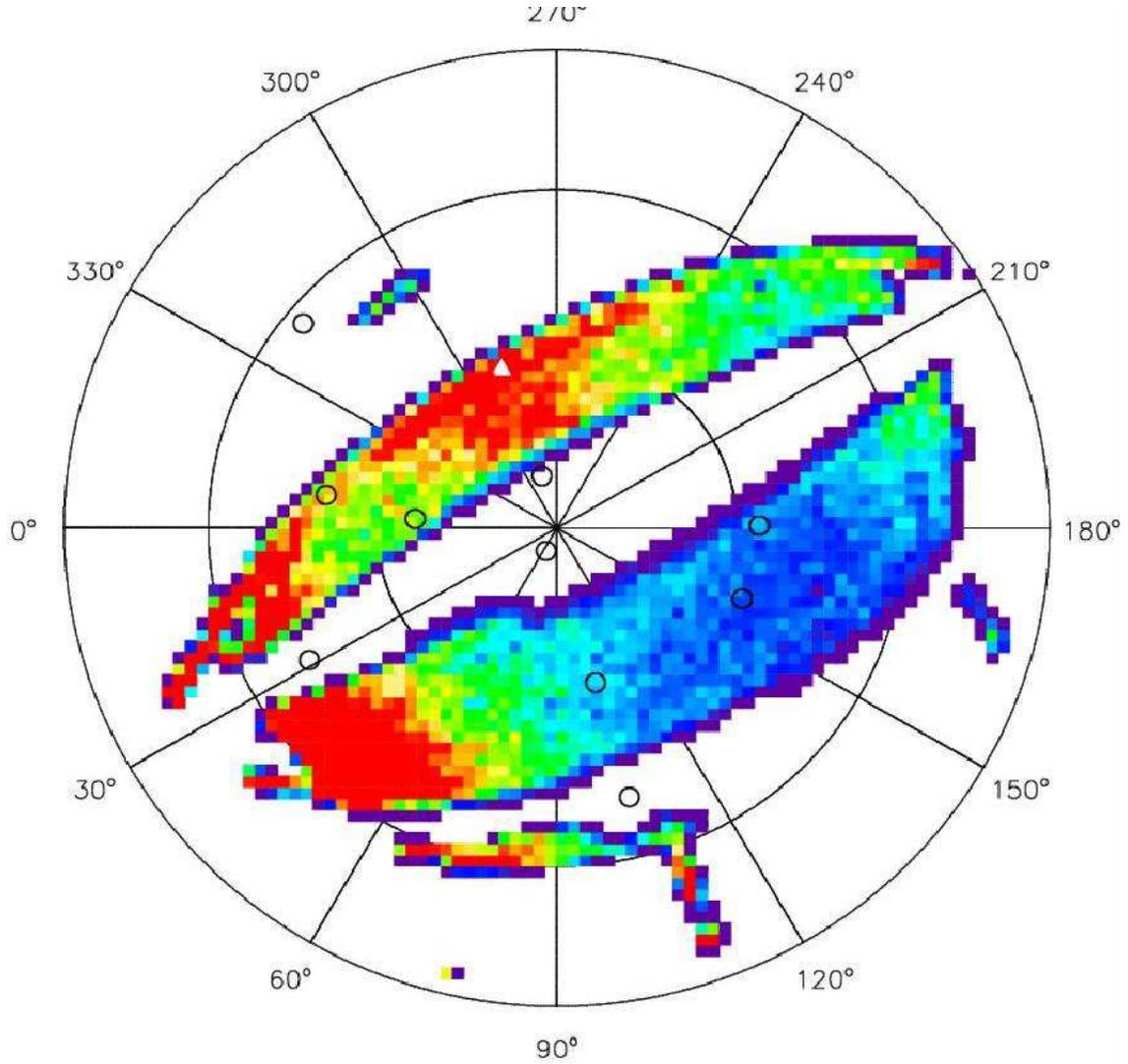}
\caption{ \small Density maps from SDSS of the North Galactic hemisphere. Position of our positive (solid triangles) and negative (open
circles)
detection of the Virgo stellar over-density given in Table 1, overplotted on a Lambert projection of
the density map of Virgo given in Juri\'c et al. (2005). This map was generated by selecting MS
stars with $b<0\deg$, $0.2 < g-r < 0.3$ and $20< r < 21$. 
Concentric circles correspond to constant galactic latitude (the North Galactic pole
is in the center; and the outermost circle correspond to b=0$\deg$. The high density patch at l=45, $b>30$ is the Milky Way disk protruding into the survey region.}

\label{densitymap}
\end{figure}

\begin{figure}
\epsscale{1.0}
\vspace{15cm}
\includegraphics{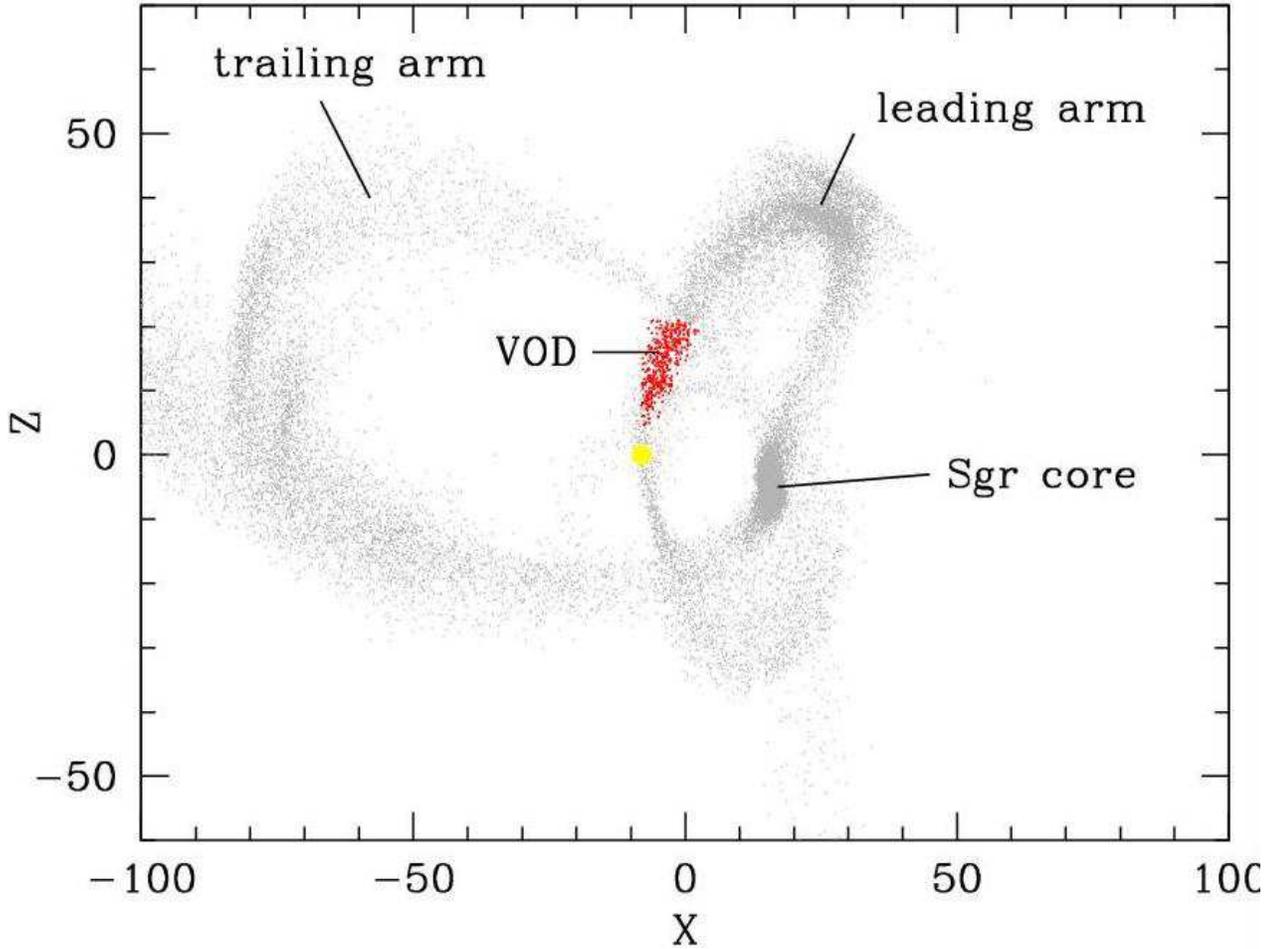}
\caption{\small Galactic  X-Z projection of  our N-body simulation of the Sgr tidal stream in a Milky Way dark matter halo with $q_{h}=0.8$. For simplicity, only particles unbound less than 2.5 Gyr ago have
been plotted. It displays the Sgr leading arm falling onto the Galactic disk, crossing the proximity of the Sun (yellow circle). The particles with a position consistent with that of the VOD are marked with red colour.}
\label{esquema}

\end{figure}

In Fig.~\ref{lambert} we plot two equal-area projections of
stream particles from the models with oblate $q_{h} = 0.8$
and prolate $q_{h} = 1.4$ halos (top and bottom, respectively).  In these plots, the North Galactic pole is in the center, concentric circles are lines of
constant Galactic latitude $b$ and the radial lines denote locations of
constant Galactic longitude $l$. The Galactic center is to the left. We restrict
the theoretical sample to the distance range of Juri\'c et al. (2006) data by
including only particles at $b > 0$ and $5 \mathrm{kpc} < d <
15 \mathrm{kpc}$. To estimate the two-dimensional projected density
distribution, we convolve the particles with a 2D Gaussian
$\sigma = 2.9\deg$ kernel. The resulting density is shown as
varying shades of gray (black and white regions denoting the highest lowest density, respectively) and can be directly compared with the observed morphology of the VOD displayed in Fig ~\ref{densitymap} (that was made with the same
scale). 

These maps show a conspicuous, diffuse lump of Sgr stream stars that extends
over a few hundreds of square degrees, whose shape and extent resembles
those of the stellar over-density displayed in Fig.~\ref{densitymap}. The
position of the over-density center (defined as the location of
maximum density) in the sky depends on the shape of the halo, and is consistent
with simulations of Law et al. (2005), Fig.~13. The oblate model
(q$_{h}$=0.8) produces a better agreement with the SDSS data, with the
over-density being centered at (l,b)=(277,63). This corresponds to a small
offset, $\sim 11\deg$, with respect to the observed position of the VOD
($(l,b)$=(300,65)) as reported by Juri\'c et al. (2006). Taking
into account the uncertainties in the Sgr stream modeling (see
Sec.~\ref{model}) and in the SDSS data as noted above, the agreement in
position is remarkable.  This result
also provides a natural explanation for the huge apparent extension of this
tidal remnant, since we are observing through a stellar tidal stream that
approaches the Sun in an almost radial direction in a localized area of the
sky.\footnote {An analogy can be found when a meteoroid stream crosses the Earth; see Nakamura et al. 2000.}

The prolate (q$_{h}$=1.4) halo model (bottom panel) produces a slightly worse 
estimate of the location of VOD center and its general shape. It does predicts a
substantial amount of faint, diffuse debris in the $180 < l < 240$ region. This is to a degree
observed in SDSS data -- e.g. the density in $180 < l < 210$ (see Fig.~\ref{densitymap}) is
apparently greater than that in its symmetric counterpart with respect to the $l=0, l=180$
direction.

Interestingly, Gondolo et al. (2005) estimated a local
stellar density of Sgr leading tail of between 270-740 stars/kpc$^{3}$,
corresponding to a range of 2.1 to 7.4 $\times$ 10$^{-7}$ stars/pc$^{3}$ or
-15.4 $< ln (\delta)<$ -14.1. This is within the measured limits of stellar
density in the VOD (see Fig. 23 in Juri\'c et al. 2006).

We note that the interpretation of the faint density excess is muddied by the existence of a different narrow
stream in that region (Juri\'c et al. 2006), apparently unrelated to
Sgr stream. An alternative scenario is that the narrow stream might be the true origin of the density excess seen in SDSS
data, instead of the Sgr debris. However, as it will become evident in the next Section, the three-dimensional structure of the VOD strongly disfavors this hypothesis.

\subsubsection{Three-dimensional spatial structure}\label{3dmap}

We compare now the three-dimensional structure of the VOD to the structure obtained from the theoretical models of the Sgr
stream in that region of the sky. We examine various cross-sections (horizontal and perpendicular to the
Galactic plane) of the VOD and the equivalent cross-sections of the Sgr
stream model.

In Fig.~\ref{xy} we plot the measured stellar density 
(left) and that obtained from the Sgr
stream models for two different values of halo flattening, $q_{h}$=0.8
(middle) and $q_{h}$=1.4 (right). 
The density shown on the top left panel is that shown in Fig.~8 (middle-left panel) of Juri\'c et al (2006). 
Before discussing its
implications, we will very  briefly describe the method used here (for a more detailed discussion we refer the reader to Juri\'c
et. al 2006): The SDSS has observed $\sim 4.2$ million G stars ($0.1 < r-i <
0.15$) within the survey area. To convert the SDSS star counts to spatial number
density, we estimated the distances to individual stars using the
photometric parallax relation of Juri\'c et al. (Eq.~$1$). From the
estimated absolute and measured apparent magnitudes, and from
astrometry, we calculate the three-dimensional $(X,Y,Z)$, positions of all stars in the
SDSS stellar sample. We then bin them to $dx=500$pc pixels in the
coordinate system as defined above to obtain a three-dimensional
star-count map. Finally, by dividing the map by the volume that was observed within each voxel, we
obtain a three dimensional map of number density within the Milky Way. 

The density shown in the left panel of Fig. \ref{xy} is a planar
cross-section of the 3D map along a plane parallel with the Galactic plane, and
at $Z=10$kpc above it. The colors encode the stellar number density (stars
pc$^{-3}$) on a natural log scale, ranging from $\exp(-15.46) = 1.9 10^{-7}$
(blue) to $\exp(-14.03) = 8.3 10^{-7}$ stars pc$^{-3}$ (red). The black regions
are places where SDSS has not observed. The yellow dot marks the position of
the Sun in X-Y plane ($X = 8$kpc, $Y = 0$kpc) The circles visualize the
presumed axial symmetry of the Galaxy. If the Galactic number density distribution was axially symmetric,
the color contours in the observed regions would trace such circles around
the Galactic center, and the density distribution would respect the symmetry
 with respect to reflection along the $X$ axis ($\rho(X,Y,Z) = \rho(-X,Y,Z)$).
As seen in the left panel, the density distribution is neither
circular nor symmetric, and exhibits a strong density enhancement centered
approximately at $X = 6, Y = 4$kpc, the VOD (Juri\'c et al. 2006).

We test under which conditions the VOD might be plausibly identified with the
leading arm of the Sgr tidal stream by over-plotting the equivalent X-Y slices at $Z=10$kpc of simulated Sgr debris over the contours of the SDSS observed
area (the gray area in middle and left panels).  The figures in the middle
and left panels are constructed in a manner similar to the method used for
SDSS data outlined above. However, we note that the particle counts have been 
averaged in the 9 kpc $ < Z <$ 11 kpc slice. The
difference in thickness, compared to $dx=500$pc thickness used for SDSS
density estimation is to account for the effect of radial smearing introduced by the intrinsic
$\sigma_{M_r} \sim 0.3$mag dispersion of the photometric parallax relation.
This effect increases the effective radial SDSS voxel scale (for a more
detailed discussion, see Juri\'c et al. 2006). We estimate the density of
Sgr debris by convolving the simulation particles in the observed
volume with a $\sigma=0.5$ kpc Gaussian kernel, thus obtaining the final
distribution shown in the panels. The color scheme is that used in
the right panel, with equal-density contours added. We also plot, as small
yellow crosses, the individual simulation particles that fall off of the
lowest equal-density contour.

\begin{figure}
\epsscale{1.0}
\vspace{15cm}
\includegraphics{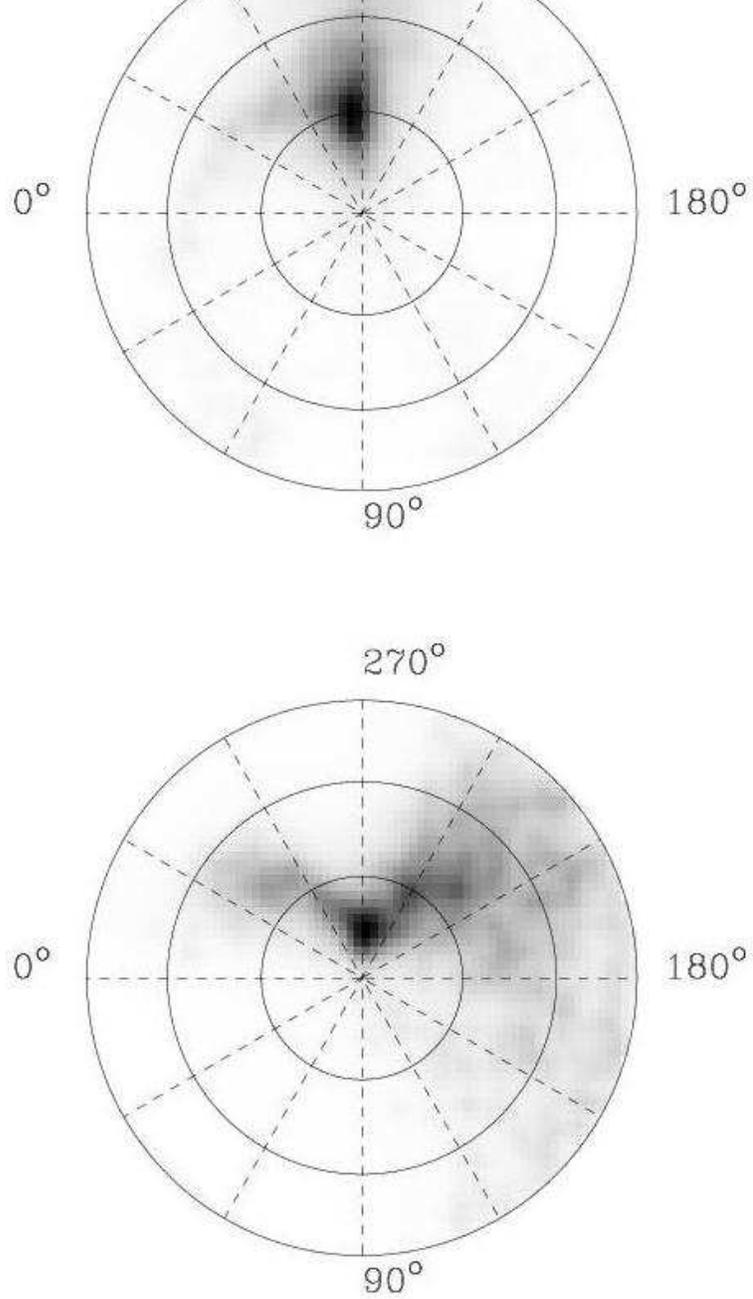}
\caption{\small Density maps of the Sgr leading tail stars with heliocentric distance 
$d_{<15 kpc}$ from our N-body simulation of the Sgr tidal stream in a Milky Way dark
matter halo with $q_{h}=0.8$ (top panel) and$q_{h}$ =1.4 (bottom panel).}
\label{lambert}

\end{figure}

\begin{figure}
\vspace{18cm}
\includegraphics{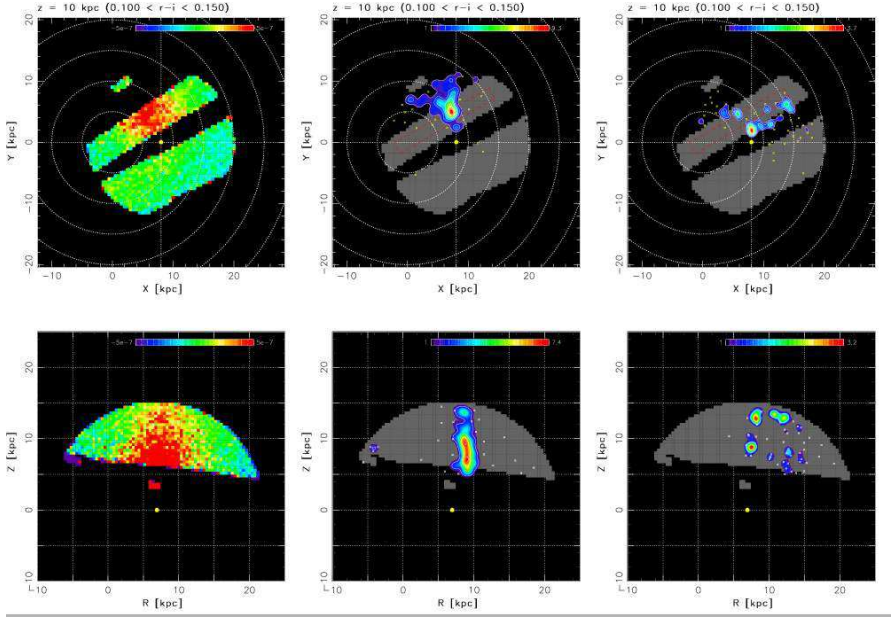}
\caption{\small  Top row: comparison of Galactic X-Y projections of the Virgo
over-density obtained from SDSS data (left column), with two Sgr stream
models, $q_h = 0.8$ (center) and $q_h = 1.4$ (right column). In the left
column, we plot the residual stellar density after subtraction of the
Galactic stellar density model in a thin $\Delta_Z = 0.5kpc$ slice at $Z
= 10$~kpc above the Galactic plane. The color coding represents the
stellar overdensity of $0.1 < r-i < 0.15$ stars in stars pc$^{-3}$. The
colored contours on the center and right rows are the Gaussian-smoothed
($\sigma=0.5$~kpc) particles from the Sgr simulation (in arbitrary
density units). To account for the observational errors, the simulation
particles were additionally scattered in the radial direction, in
accordance with the measured SDSS photometric scatter and the intrinsic
scatter of the photometric paralax relation (Fig 4., Juric et al. 2006).
Bottom row: comparison of stellar number density in SDSS slice of Virgo
over-density with the Sgr simulation. The $Z$ axis denotes the distance
from the Galactic plane. The $R$ coordinate is the distance from the
Galactic center in the $\phi=30^O$ direction counter-clockwise along the
axis of Galactic rotation. The color coding and the symbols are the same
as in the top row.}
\label{xy}
\end{figure}

As it was the case in Fig.~\ref{lambert}, the oblate $q_h = 0.8$ model agrees well with the data
within the observational and theoretical uncertainties. It successfully reproduces the diffuse shape
of the over-density and the location of over-density maximum. The location of the maximum for $q_h =
0.8$ model locates within 1.5 to 2kpc of the observed $R \sim 6.5-7$kpc maximum of VOD, and the
diffuse morphology of the $q_h = 0.8$ model appears to show a considerably better match to the observed
morphology than that of the $q_h = 1.4$ model. Some debris from the $q_h = 1.4$ model (right panels)
can be also found at the location of the VOD. However, this model predicts the existence of other
over-densities where none have been detected and the morphology at the VOD location is irreconcilable 
with that inferred from the SDSS data.

We now examine the vertical distribution of matter in the VOD and compare it with the models of Sgr stream. The bottom panels of Fig.~\ref{xy} compare the observed vertical stellar number
density distribution in the Galaxy, cut along a narrow $\phi=30\deg$ plane
that passes through the center of the Galaxy and the center of the VOD (the red dashed
box on top middle and left panels of Fig.~\ref{xy}). The horizontal coordinate is the distance from
the Galactic center, $R$, along the $\phi=30\arcdeg$ direction, while the vertical
coordinate is the height above the Galactic plane. As in the case of the top left panel, the
bottom left panel shows the over-density as measured from SDSS data (and is equivalent to the top
right panel of Fig.~22 in Juri\'c et al.2006), while the two panels to the right show the simulated
density of the Sgr tidal stream as obtained from our simulations (Sec.~\ref{model}) using an oblate
($q_{h}$=0.8) and a prolate ($q_{h}$=1.4) halo. The color scheme is the same as for the top panels,
except that in the case of the bottom left panel we show the fractional residuals with respect to
the best fit Galactic halo model of Juri\'c et al. (2006).\\
In the case of $q_{h}$=0.8 halo, the Sgr leading arm is nearly perpendicular to the
Galactic plane, with a slight downward density gradient, in agreement with the findings of Juri\'c
et al. (2006). Its location agrees to within $\Delta R \sim 1$ kpc of the observed position of Virgo over-density. In contrast, although the prolate models 
($q_{h}$=1.4)  correctly reproduce the Galactocentric distance of the over-density ($R \sim 6-7$kpc), they fail in order to reproduce the observed morphology. These models show a strong density maximum
at $(R,Z)\simeq (7,8)$kpc (incompatible with the density map) and pieces of debris scattered out to $R \sim 12.5$kpc (not detected by SDSS). In this working scenario, the VOD morphology strongly disfavours prolate halo models.

The conclusions obtained from these cross-sections can be summarized in the different spatial perspectives displayed in Figure \ref{3dmap}\footnote{This figure is a snapshot of a set of movies included as supplementary material  in the electronic version of this paper and also available at
http://www.astro.princeton.edu/$\sim$mjuric/sgr-virgo/}, which clearly shows  the coincidence in position
of the VOD and the location of the Sgr leading tail. Note that this part of the stream distributes nearly perpendicularly
 to the Galactic disk in this direction of the sky. The panels of Fig.~\ref{3dmap} 
show  two different 3-dimensional
 projections (top and bottom row) of our Sgr model computed for oblate 
$q_{h}=0.8$ (left column) and prolate $q_{h}=1.4$ halos (right column). They
visualize the simulated Sgr dwarf and its tidally stripped
tails in rainbow color map, with the color encoding the column density of
simulation particles (red being the highest, blue the lowest), which are over-plotted
in the panels as black dots. Note that the core of Sgr dwarf corresponds to
the darkest region owing to the high density of N-body particles
that are still gravitationally bound to the dwarf's remnants. The
coordinate system origin is located at the Galactic center with the red axis (X) pointing
towards the Sun (yellow circle) and the blue axis (Z) pointing towards the
North Galactic pole.
The region in which Juri\'c et al. (2006) detect the VOD is
marked by the white wire-frame box, whereas the location of VSS as reported by Duffau et al. (2006; see Sec. 1) is shown by the blue wire frame.
 This coincidence points to a
plausible identification of VOD with the leading arm of the Sgr tidal
stream or, at the very least, to likely significant pollution of VOD as seen in the SDSS density maps by the Sgr tidal debris. 

 Taken together, the analysis of figures~\ref{densitymap}, \ref{lambert}, \ref{xy} and ~\ref{3dmap} clearly suggests that the Virgo over-density may correspond to the detection of the Sgr stream's leading arm falling onto the Milky Way plane. Furthermore, the observed morphology is best described by models assuming an {\em oblate} Milky Way halo.   
With these results at hand, it appears fairly unlikely that the VOD is a tidal stream unrelated to the Sgr dwarf and yet it has the predicted location and morphology of the Sgr leading arm.

\subsubsection{Radial velocity of the Virgo over-density}\label{radialveloc}

At present there is no kinematic information available on the stars of Virgo over-density. Radial velocities are necessary in order to establish unambiguously whether the Virgo over-density is associated to the Sgr stream. Theoretical models predict that, independently of the halo shape, the leading arm of the Sgr stream is present in this area of the sky and moves on average with negative radial velocities as it {\it falls} onto the Milky Way disk. However, the exact magnitude of the mean radial velocity is a vital discriminator for the halo shape, as prolate and oblate halos lead to fairly distinct trends of $V_r$ along the leading arm (Mart\'inez-Delgado et al. 2004a; Helmi 2004, Law et al. 2005).
We must remark that at present there is no information on the kinematics of the Sgr leading arm in this area of the sky, since this part of the stream  was not covered by the spectroscopic survey of M-giants carried out by Majewski et al. (2004).

There have been radial velocity measurements of RR Lyrae stars in the Virgo constellation that have shown the presence of a spatial over-density (called the {\it 12$^{h}$.4 clump}, see Sec. 1) located at $D\simeq 20$ kpc. 
Duffau et al. (2006) find a mean radial velocity in the galactic rest frame of $(V_r)_{gsr}\simeq 99.8$ km/s (see Table 1) with a fairly low velocity dispersion ( $\sigma$=17.3 km s$^{-1}$), unexpected for the smooth population of RR Lyrae in the Milky Way, that suggests a possible tidal stream detection. According to the theoretical models of the Sgr stream, the positive value of  $(V_r)_{gsr}$ excludes a membership of these stars to the leading arm. 
However, it is unclear whether the {\it 12$^{h}$.4 clump} is not associated to the Sgr stream at all, since the oblate halo models ($q_h=0.8$ in this work) also predict the presence of the trailing arm stars in that volume of the Milky Way (see Fig.~\ref{3dmap}, where the blue box represents the {\it 12$^{h}$.4 clump} position) with an averaged radial velocity of $(V_r)_{gsr}$= 90 km/s, fairly similar to that reported by Duffau et al. (2006) ($(V_r)_{gsr}$=+83 or +100  km/s, depending of their sample selection). Therefore, if the  {\it 12$^{h}$.4 clump} proves to belong to the Sgr trailing arm it would favor the models with $q_h<1$.

\section{DISCUSSION}\label{discussion}
Theoretical models of the Sgr stream are not able to reproduce all the observational constraints. All models agree that the geometry of the tidal stream (i.e its 3D spatial distribution) clearly rules out halo models with a prolate shape (Mart\'inez-Delgado et al. 2004a, Johnston et al. 2005, Law et al. 2005) and that Milky Way halos with $q_h<1$  are preferred at a $3\sigma$ level.
Paradoxically, the models that best explain the stream structure fail to reproduce the radial velocity trend along the leading arm unless prolate halo models ($q_h>1$) are invoked (Helmi 2004). Apparently, the present status of the Sgr stream modeling is that of a fish biting its tail.

It is not clear how to reconcile observations and theory. Possible explanations for this mismatch are: (i) the Milky Way models are over-simplistic (although similar models successfully reproduce other tidal streams such as Monoceros, Pe\~narrubia et al 2005, and the Magellanic stream,  Ruzicka, Palous \& Theis 2006, both streams being best reproduced by models with oblate halos) or (ii) the presence of possible systematic errors on the data of the Sgr tidal stream used in
the computation of these simulations (as suggested by Chou et al. 2006).

In this contribution, we have investigated the origin of the Virgo over-density discovered from the analysis of the SDSS data. The VOD is the largest clump of tidal debris ever detected in the outer halo of the Milky Way. We find that the position, apparent angular size, stellar density  and three dimensional geometry match the theoretical predictions from the present Sgr stream models. According to these models, the Virgo over-density corresponds to  Sgr leading arm falling onto the Milky Way disk. In agreement with other portions of the Sgr stream, the spatial distribution of the VOD is best reproduced by models with oblate halos $q_h<1$. However, we must remark that a unambiguous association between the VOD and the Sgr stream cannot be established in absence of stellar kinematics.

If the Virgo over-density proves to belong to the Sgr stream it might represent the key piece to solve the paradox described above, as we have for the first time the opportunity to measure the radial velocity of {\em millions} of stream stars just in the region where observations and theoretical models do not match. 

Additionally, the detection of a new portion of the stream's leading arm provides strong constraints on future theoretical models and will help to address whether the Sgr stream passes through the solar neighbourhood as it crosses the Milky Way 
plane (as predicted by  Kundu et al. 2002, Majewski et al. 2003; Mart\'\i nez-Delgado et al. 2004). In that case, the Sgr stream would represent an important target for the on-going RAVE survey (Steinmetz et al. 2006), whose goal is to measure the radial velocity of around $10^6$ stars in both hemispheres. Thanks to the large sample of radial velocities, if the Milky Way crossing occurs at a heliocentric distance $D< 3$ kpc, its presence will be detected in form of an excess of stars with $v_r<0$ in the Northern hemisphere ($b\geq 0$) and with $v_r>0$ for $b<0$ with respect to the background population of Milky Way stars (Pe\~narrubia, Navarro \& Helmi, in preparation). 
The presence of a tidal stream in the solar neighbourhood also represents an excellent target for Weakly Interacting Massive Particles (WIMPs) direct detection experiments, since a substructure of dark matter in
the proximity of the Sun would yield a ``cold'' flow of WIMPs thought the
detector of either of these dark matter experiments (see Savage, Freese and
Gondolo 2006 and Gondolo et al. 2005 for details). This would make the solar vicinity an excellent lab for probing the physical properties and nature of the dark matter.

\vskip1cm

We thanks to Brian Yanny and Kathy Vivas for making available to us some resultson the Virgo tidal stream  during the preparation of this paper. We also thanks
to the anonymous referee for useful comments that helped to improve this manuscript.
DMD acknowledges funding by the Spanish Ministry
of Education and Science (Ramon y Cajal contract and research project AYA 2001-3939-C03-01). JP thanks Julio Navarro for financial support.

   Funding for the SDSS and SDSS-II has been provided by the Alfred P.
Sloan Foundation, the Participating Institutions, the National Science
Foundation, the U.S. Department of Energy, the National Aeronautics and
Space Administration, the Japanese Monbukagakusho, the Max Planck
Society, and the Higher Education Funding Council for England. The SDSS
Web Site is http://www.sdss.org/.

    The SDSS is managed by the Astrophysical Research Consortium for the
Participating Institutions. The Participating Institutions are the
American Museum of Natural History, Astrophysical Institute Potsdam,
University of Basel, Cambridge University, Case Western Reserve
University, University of Chicago, Drexel University, Fermilab, the
Institute for Advanced Study, the Japan Participation Group, Johns
Hopkins University, the Joint Institute for Nuclear Astrophysics, the
Kavli Institute for Particle Astrophysics and Cosmology, the Korean
Scientist Group, the Chinese Academy of Sciences (LAMOST), Los Alamos
National Laboratory, the Max-Planck-Institute for Astronomy (MPIA), the
Max-Planck-Institute for Astrophysics (MPA), New Mexico State
University, Ohio State University, University of Pittsburgh, University
of Portsmouth, Princeton University, the United States Naval
Observatory, and the University of Washington.

\newpage

\newpage

\begin{figure}
\epsscale{1.0}
\plotone{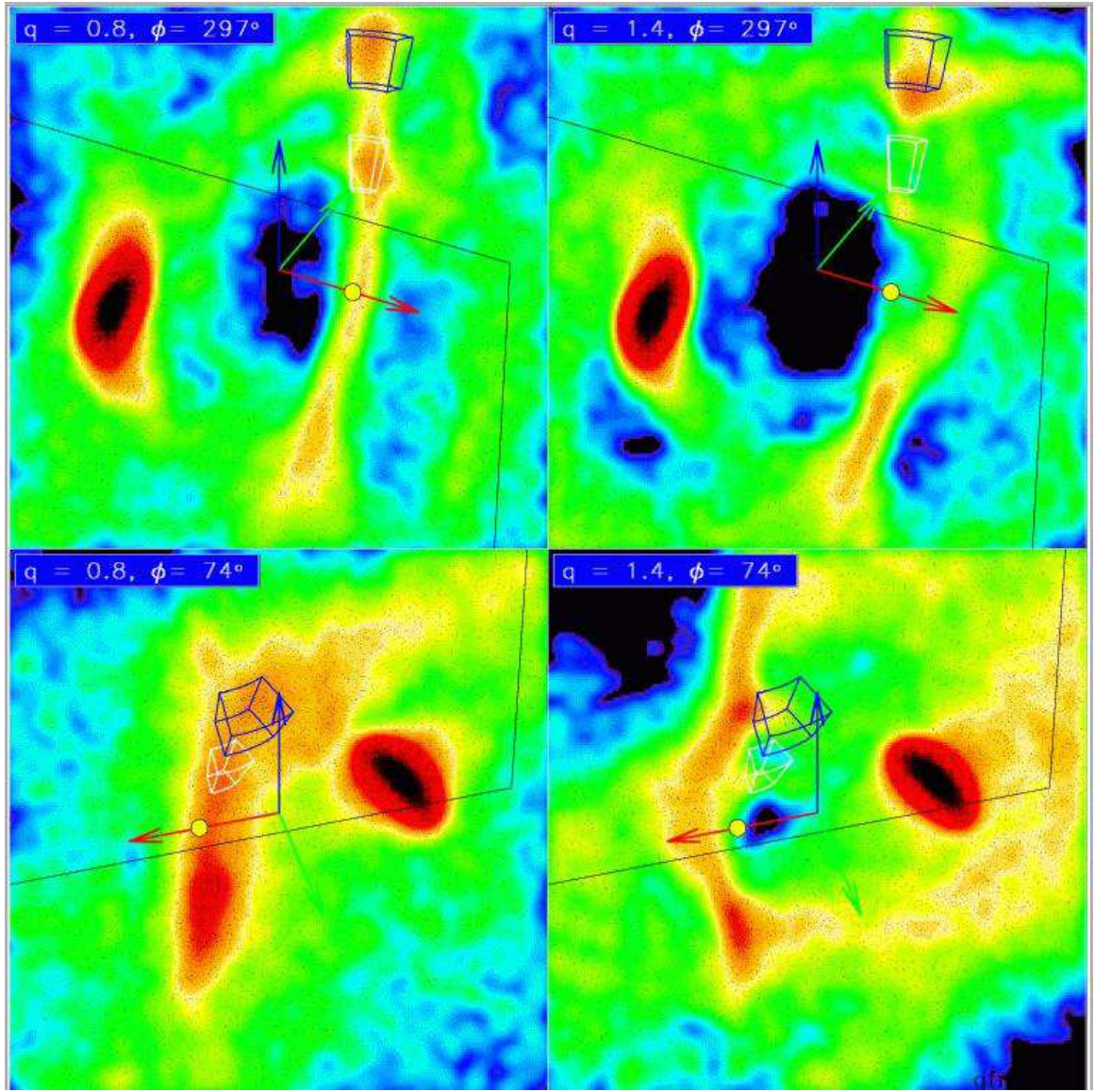}
\caption{\small Two different 3-dimensional projections (top and bottom rows)of the Sagittarius tidal stream
models computed for an oblate halo $q_{h}=0.8$ (left column) and a prolate $q_{h}=1.4$ halo (right
column). The column density of simulation particles is represented by rainbow color-map, with the red
being the areas of highest, and blue the lowest density. Over-plotted on the column density map as
black dots are the simulation particles themselves. The core of Sgr dwarf is clearly visible as a
dense, dark region, due to the high density of simulation particles which are still gravitationally
bound to the dwarf's core. The origin of the coordinate system axes is the Galactic center, with the
red axis (X) pointing
towards the Sun (yellow circle), the blue axis (Z) pointing towards the north Galactic pole.
The region in which Juri\'c et al. (2006) detect the Virgo overdensity is marked by the white
wire-frame box. The blue wire-frame marks the location of Virgo Stellar Stream as reported by Duffau
et al. (2006).
}.
\label{3dmap}
\end{figure}

\newpage

\begin{table}
\begin{center}
 \caption{Positive and negative detections of Virgo from colour-magnitude
diagrams \label{tbl-2}}

\vspace{0.5cm}

\begin{tabular}{lccccc}
\tableline\tableline
 RA(J2000)  & Dec(J2000) & $l$ & $b$& Detection  \\
\tableline
\tableline

   12:24:00.00 & -1:00:00.0   &  288.62 &  61.11  &  Positive  \\
   12:24:00.00 &  -0:30:00.0  &    288.39 &   61.59 &    Positive \\
   10:20:00.00 &  40:00:00.0   &     180.71 &  56.24  &    Negative \\
   13:48:16.00 &  52:08:32.0   &   103.69  & 62.81  &     Negative \\
  15:37:49.00 &   69:17:31.0  &    104.68 &  41.44   &   Negative \\
  12:40:00.00 &   18:20:00.0  &    285.67 &  80.82  &    Negative \\
  13:56:21.00 &  -27:10:04.0  &    320.28 &  33.51   &   Negative \\
  13:07:53.00 &   29:21:00.0  &    65.42 &  85.75   &   Negative \\
  10:49:22.00 &   51:03:04.0  &   158.57 &  56.78   &   Negative \\
  14:07:37.00 &   11:50:28.0  &     356.10 &  66.48   &   Negative \\
  16:11:05.00 &   14:57:29.0  &     28.76 &  42.18   &   Negative \\
  14:44:27.00 &  -1:00:00.0  &    351.54 &  50.88  &     Negative \\

\tableline

\tableline
\end{tabular}

\end{center}
\end{table}

\end{document}